\documentclass[fleqn,usenatbib]{mnras}

\usepackage{newtxtext,newtxmath}

\usepackage[T1]{fontenc}
\usepackage{ae,aecompl}

\usepackage{pifont}
\usepackage{natbib}
\usepackage{fleqn}

\usepackage{amsmath}
\usepackage{amssymb}
\usepackage{bm}

\usepackage{graphicx}
\usepackage{caption}
\usepackage{subcaption}
\usepackage{cleveref}
\usepackage{multirow}

\title[An Algorithm for the Visualization of Relevant Patterns in Astronomical Light Curves]{An Algorithm for the Visualization of Relevant Patterns in Astronomical Light Curves}

\author[C. Pieringer et al.]{
Christian Pieringer,$^{1}$\thanks{E-mail: cpieringer@inacap.cl or cppierin@uc.cl}
Karim Pichara,$^{2,5}$\thanks{E-mail: kpb@ing.puc.cl}
Marcio Catelan$^{3,5}$\thanks{E-mail: mcatelan@astro.puc.cl}
Pavlos Protopapas$^{4}$\thanks{E-mail: pavlos@seas.harvard.edu}
\\
$^{1}$Universidad Tecnol\'ogica de Chile -- INACAP. Santiago, Chile.\\
$^{2}$Department of Computer Science. School of Engineering, Pontificia Universidad Cat\'olica de Chile. Santiago, Chile.\\
$^{3}$Institute of Astrophysics. Pontificia Universidad Cat\'olica de Chile. Santiago, Chile.\\
$^{4}$Institute for Applied Computational Science (IACS). Harvard University. Cambridge, MA. USA.\\
$^{5}$Millennium Institute of Astrophysics. Santiago. Chile.\\
}

\date{Accepted XXX. Received YYY; in original form ZZZ}

\pubyear{2018}

\begin{document}
\label{firstpage}
\pagerange{\pageref{firstpage}--\pageref{lastpage}}
\maketitle

\begin{abstract}
Within the last years, the classification of variable stars with Machine Learning has become a mainstream area of research. Recently, visualization of time series is attracting more attention in data science as a tool to visually help scientists to recognize significant patterns in complex dynamics. Within the Machine Learning literature, dictionary-based methods have been widely used to encode relevant parts of image data. These methods intrinsically assign a degree of importance to patches in pictures, according to their contribution in the image reconstruction. Inspired by dictionary-based techniques, we present an approach that naturally provides the visualization of salient parts in astronomical light curves, making the analogy between image patches and relevant pieces in time series. Our approach encodes the most meaningful patterns such that we can approximately reconstruct light curves by just using the encoded information. We test our method in light curves from the OGLE-III and StarLight databases. Our results show that the proposed model delivers an automatic and intuitive visualization of relevant light curve parts, such as local peaks and drops in magnitude.

\end{abstract}




 \section[Introduction]{Introduction}

In latest years, the modern astronomical hardware has made it possible to access and collect comprehensive information about the sky, switching the way that we understand the universe \citep{wootten2003atacama,dewdney2009square,thompson2013real,LSSTbook}. The massive amounts of data cannot be stored and analyzed in the way that we are used to because the current storage capacity surpasses our ability to process it. In this sense, machine learning is an essential piece and powerful tool for analyzing and extracting new knowledge from massive, high-dimensional and noisy datasets. Machine learning has contributed remarkably to time domain analysis allowing for the understanding of astrophysical phenomena providing valuable insight into astronomical objects that change through time \citep{debosscher2007automated,richards2011machine,bloom2012automating,pichara2012improved,graff2013skynet,pichara2013automatic,nun:pichara:2014,mackenzie2016clustering,pichara2016meta}.

Light curves are a time series that encode the light variations of a pulsating object along the time. Automatic methods applied for light curves classification train a predictive model using templates or known type of variability (labels). On the one hand, the use of templates assumes that each object in a survey can be well modeled at least by one instance in the templates. This assumption may be unreasonable for large datasets and may cause a poor classification \citep{richards2012semi}. On the other hand, the use of labeled dataset requires a cross-matching procedure to merge two astronomical catalogs. During cross-identification, we typically rely on the Right Ascension (RA) and the Declination (DEC) sky coordinates of known objects in one catalog to assign the object label for the closest object in the other catalog, assuming that the systematic positioning error between the two catalogs is zero \citep{nieto2007cross}. Cross-matching is hard to solve because of the high computational complexity of large-scale catalogs, and the distance error between objects \citep{nieto2007cross,du2014new}.

Furthermore, the matching procedure has significant problems in very crowded fields, such as the Galactic plane. This factor has implications on merging catalogs because matching algorithms will miss some matches due to significant flux contamination by faint and close objects, and finally carrying labeling mistakes for new catalogs \citep{wilson2017effect}. These facts suggest that in some cases it may be reasonable to create a training dataset from scratch.

Data visualization is another useful set of tools in the data science pipeline that aims to present and to encode the information precisely and efficiently as a visual object in a more informative and human-readable way. Visualization has become relevant in various domains \citep{fu2008stock,Tao2016}. For example, in the astronomical context, a folded light curve of a variable star presents typical visual local patterns that we might discriminate by a set of salient waveforms. Recognizing these parts is essential for a human visual identification process and an enhanced analysis \citep{li2013time}. The automatic detection and visualization of relevant local parts based on machine learning algorithms has a high impact on communicating the information contained in an astronomical time series.

Dictionary-based classifiers use a dictionary of prototypes to encode an input signal as a linear combination of these prototypes. These methods have shown to be completely useful detecting and recognizing local patterns in other domains such as computer vision \citep{wright2010sparse,mairal2014sparse}, health signal \citep{wang2013human,yuan2014epileptic}, and audio signals \citep{ness2012auditory,zubair2013dictionary}. Its advantage falls on the fact that these methods take into account the local effects of prototypes in each part of the data. In this sense, we could able to identify and visualize where the model puts more efforts during reconstruction or in a classification model. 

We propose a framework for providing to any scientist with visual clues where they have to pay attention to the time series in astronomical data sets. Our method learns the salient parts in light curves and provides help during the visual identification.

\begin{enumerate}
    \item the use of a sparse approximation allows us to process time series providing a smooth and highlighted version of the real pattern in the data without using regression,
    \item the set of encoding vectors provides an embedded mechanism to highlight salient parts in light curves,
    \item the detection of salient parts allows us to organize an observational schedule for variable objects.
\end{enumerate}

The article describes our method and results in synthetic and real datasets. Results show that our framework can be used promisingly in visual identification of astronomical time series and for observational planning. The rest of the text organizes as follow. Section 2 includes related work to visualization. Section 3 presents our algorithm and its theoretical foundations. Section 4 describes the datasets, the experiments and the results that we collect. Finally, Section 5 contains the conclusions and closing remarks from our approach. 

\section{Dictionary-based Learning}
\label{sec:dictionary_based}

Dictionary-based learning is a state-of-art methodology used for approximating an input as a linear combination of an over-complete set of prototypes or atoms \citep{murphy2012machine}. These prototypes represent most of the variations of the input signal, and it would be useful for reconstructing any class instance in the dataset. A decoding algorithm uses the dictionary to fit a new version of the light curve using a regularized linear combination of the atoms independently of the class from they where taken \citep{mairal2009supervised,olshausen1997sparse}.

A particular case of this learning algorithms family is Sparse Coding (SC). It is an unsupervised method that aims to build a dictionary of atoms able to encode and reconstruct a signal. A Sparse Coding algorithm typically learns the dictionary of representative atoms in a two-step optimization process using an unsupervised training set. First, the algorithm finds a set of atoms to represent the data. Second, it finds via a non-linear encoding scheme the linear set of estimators for representing the input signal according to the dictionary.  The optimization ends when the reconstruction error is minimum. 
In our context of astronomical time series, each atom in the dictionary would represent a waveform prototype of the morphology of these data.

This section provides a theoretical explanation of sparse-based reconstruction and the stages involved in building the dictionary.

\subsection{Sparse-based Reconstruction}
\label{sec:SparseBasedReconstruction}


Sparse-based methods rely on the assumption that any portion of an input signal can be approximated using a few sets of elements in a dictionary of representative parts of it. Elements in the dictionary receive the name of atoms. The approximated signal is a smooth version of the original signal and in some context reducing the noise of the input \citep{mairal2009supervised}. These methods achieve the approximation by encoding the input through a linear combination of the atoms.


Let $\mathbf{X} = [\mathbf{x}^1, \mathbf{x}^2, \ldots, \mathbf{x}^N]^T \in \mathbb{R}^{M \times N}$ be a set of $N$ input signals $\mathbf{x}^i \in \mathbb{R}^M$, $\mathbf{D} = [\mathbf{d}^1, \mathbf{d}^2, \ldots, \mathbf{d}^K] \in \mathbb{R}^{M \times K}$ be an over-complete dictionary with $K$ atoms $\mathbf{d}^j \in \mathbb{R}^M$, and $\mathbf{A} = [\bm{\alpha}^1, \bm{\alpha}^2, \ldots, \bm{\alpha}^K] \in \bm{R}^{M \times K}$ be a matrix of coding vectors $\bm{\alpha}^k$. A sparse-based algorithm minimizes the reconstruction error of the input signal $\mathbf{X}$ as follows:

\begin{eqnarray}
	\label{eq:SCAproximation}
    \underset{\mathbf{A}}{\text{minimize}} \quad \|\mathbf{X} - \mathbf{DA}\|_{F}^{2} + \lambda\|\mathbf{A}\|_{\ell_r},
\end{eqnarray}

\noindent
where 
$\lambda$ is a regularization parameter. The $\ell_r$ penalty provides a sparse solution for $\bm{\alpha}^k$. In case $r = 0$ the solution fits the Orthogonal Matching Pursuit (OMP) algorithm \citep{pati1993orthogonal}. The OMP algorithm handles this optimization problem using a greedy approach. In this case, the algorithm finds the best matching atoms iteratively in the dictionary that minimize the reconstruction error in the sense of the $L_0$ norm. In cases where $r = 1$ the minimization matches the solution proposed for SC \citep{olshausen1997sparse}. The algorithm finds the solution according to Lasso regression \citep{tibshirani1996regression}. This regression method allows variables selection and regularization, enhancing the prediction accuracy of the model it produces.



%

%

\subsection{Dictionary Learning}


As we explain in section \ref{sec:SparseBasedReconstruction}, sparse-based models learn the data representation via a nonlinear encoding scheme. Also, these methods can simultaneously find the dictionary with the best atoms that fit specific data. The way of solving this problem is to alternate between the two variables $\mathbf{D}$ and $\mathbf{A}$, minimizing over one while keeping the other one fixed in the next optimization problem \citep{mairal2009supervised,olshausen1997sparse}:



\begin{equation}
    \underset{\mathbf{D,A}}{\text{minimize}} \quad \|\mathbf{X} - \mathbf{DA}\|_{F}^{2} + \lambda\|\mathbf{A}\|_{\ell_r}, \label{SPDictionary}
\end{equation}

As \citet{coates2011importance} conclude, the main advantage of sparse-based models is not learning better atoms but arises from its non-linear encoding scheme. Furthermore, SC solves a convex optimization problem iteratively that finally could be very expensive to deploy with large amounts of data. In this sense, the authors propose recommendations and conclusions about using \textit{K-Means} to design the dictionary. \textit{K-means} is a well-studied method to learn features directly from raw inputs in various domains. It is a simple, fast and scalable algorithm. \textit{K-means} build the dictionary solving the following problem:

\begin{equation}
   \begin{aligned}
   & \underset{\mathbf{D},\bm{\alpha}}{\text{minimize}} & & \sum_{i}\|\mathbf{x} - \mathbf{D\alpha}_i\|_{2}^{2}\\
   & \text{subject to} & & \|\mathbf{\alpha}_i\|_{0} \leq 1, \forall_{i}, \label{SPKMeans}
   \end{aligned}
\end{equation}

\noindent where, similarly to SC, $\mathbf{D} \in \mathbb{R}^{m \times k}$ is the dictionary, and $\mathbf{\alpha}_i \in \mathbb{R}^{k}$ is the code vector associated with the input $\mathbf{x}$.

\section{The Approximation Method}
\label{sec:method}

In this section, we propose and describe a methodology to reconstruct astronomical time series using sparse-based models and a dictionary with atoms previously learned from light curves.

\subsection{Pre-processing}
\label{pre-processing}

Before training the dictionary, our algorithm normalizes each light curve to zero-mean and unit variance. Then, it performs a rolling window of size $M$ along every light curve $l$, where $\mathbf{L} = [\mathbf{l}^1, \mathbf{l}^2, \cdots, \mathbf{l}^R] \in \mathbb{R}^{M \times R}$ is the set of overlapped windows after this pre-processing. Finally, the algorithm randomly grabs a set of $s < R$ windows from the $r$-th light curve.

\subsection{Dictionary Learning}
\label{DictionaryLearning}

To train the dictionary, we use a mini-batch version of K-Means to handle a large amount of data \citep{scikit-learn}. First, the algorithm randomly initializes the centroids from a Normal distribution and then normalizes them to unit length. The algorithm reinitializes empty cluster centroids using random examples from sampled windows, as suggested in \citet{coates2012learning}.

In each iteration, K-Means takes a chunk of $t > s$ sampled windows from a set of light curves, computing the new cluster centers, and updating the centroids calculated in the previous iteration. We substitute the centroids by the nearest sampled window in the training chunk. This procedure allows us to build a dictionary using real atoms from the dataset. As suggested in \citep{marascu2014tristan}, we build our dictionary from a set of per-class dictionaries. This means that $\mathbf{D} = \bigcup_{j=1}^k{\mathbf{D}_j}$, where $k$ is the number of classes. This approach allows us to include all the representative atoms from each class to the final dictionary.

\subsection{Approximation Algorithm}
\label{ApproximationAlgorithm}

The approximation algorithm has two main stages: encoding and reconstruction. During the encoding stage, the algorithm performs the rolling window pre-processing along the light curve to encode. It is the same sampling procedure performed during training. Then, the algorithm encodes each window into an $\alpha_i$ vector applying the dictionary learned previously. We try two off-the-shelf encoding algorithms: OMP and LASSO. In the reconstruction, the algorithm builds a new version of the light curve from the encoded windows. This version is free of noise. We achieve the reconstruction using equation \eqref{eq:SCAproximation}, where $\mathbf{X} = [\mathbf{x}^1, \mathbf{x}^2, \ldots, \mathbf{x}^n]^T$ and $\mathbf{x}^i$ is the window after the rolling window procedure. The algorithm merges all the segments $\mathbf{x}^i$ by consecutively adding all the segments and dividing by the overlap.

Once the reconstruction is ready, the algorithm computes the relevance of each part of the light curve using the weight of the atoms provided by the encoder to each time series segment. Let $\omega_i$ be the relevance of the segment $\bm{x}^i$, we calculate the maximum weight of it as $w_i = \text{max}(|\bm{\alpha}^i|)$ such that $\bm{\omega} = \bigcup_{i=1}^{n} w_i$. Finally, we normalize $\bm{\omega}$ in $[0, 1]$. During training and the reconstruction, our framework does not consider the error or the uncertainties related to measurements in the light curve.  Our algorithm provides a visualization with a smooth version of the original light curve.

\section[Results]{Results}
\label{sec:results}

This section describes the studies we conduct to analyze the contribution of our approach. In each case, we describe the datasets and our results.

\subsection{Databases Description}
\label{DatasetDescription}

\begin{itemize}
  \item Starlight: It is a dataset from the University California Riverside (UCR) Time Series Classification Archive that combines light curves from the MACHO and OGLE surveys \citep{UCRArchive,protopapas2006finding}. The dataset contains 9,236 folded light curves through period estimation \citep{lomb76,reimann1994frequency}, and where the resulting folded curve has 1,024 data points. It has three classes of variable stars: 2,580 Eclipsing Binaries (EB), 1,329 Classical Type-I Cepheids (CEP) and 5,237 RRab and RRc RR Lyrae (RRL). Training set contains 1,000 objects and the test data 8,236 objects, distributed as shown in Table \ref{tb:StarLightDistribution}. We refer the reader to \citep{protopapas2006finding} for more details about this dataset.

\begin{table}
\centering
\caption{Distribution of classes in the StarLight data set. Most of the instances are RRL.}
 \label{tb:StarLightDistribution}
\setlength\arrayrulewidth{1pt}
  \begin{tabular}{lcc}
  	\cline{2-3}
  	& \multicolumn{2}{c}{\textbf{No. of Instances}}  \\
    \hline
  	\textbf{Class} 	&	\textbf{Training}	&	\textbf{Test} \\
    \hline
    CE		& 	152			&  1,777	\\
    EB		& 	275			&  2,305 \\
    RRL 	& 	573			&  4,754 \\
    \hline
    \textbf{Total}	&  \textbf{1,000} &  \textbf{8,236} \\
    \hline
  \end{tabular}
\end{table}

 \item OGLE-3: It collects the data from the third phase of the Optical Gravitational Lensing Experiment \citep[OGLE;][]{udalski2008optical}. It organizes into the four targets: Galactic bulge (BLG), Galactic disk (GD), Large Magellanic Cloud (LMC), and Small Magellanic Cloud (SMC). The primary survey focuses on variable objects in the I-band filter. However, for all the fields it also includes some observations in the V-band. We specifically use SMC field that includes a total number of 30,550 light curves distributed as follows: 2,768 CEP; 6,138 EB; 19,384 Long Period Variables (LPV); 2,217 RRL; and 43 Type II Cepheids (T2CEP)
\end{itemize}

\subsection{Case Study 1: Reconstruction}

In this experiment, we analyze the reconstruction properties of the proposed approach even in noisy conditions. We use the datasets described in Section \ref{DatasetDescription}. This experiment allows us to know the behavior of our algorithm and adjust parameters such as the window size, the number of atoms in the dictionary and their effects on the final reconstruction.

First, we train the dictionaries as we describe in Section \ref{DictionaryLearning} getting samples from the training dataset. We implement supervised training, where we allow the algorithm to train one dictionary per class. We use a fixed number of atoms $k=\{32, 64, 96\}$, and windows size $m=\{64, 128, 256, 384\}$. During dictionary training, we limit the algorithm to get only ten samples per light curve from 150 light curves per training chunk to feed the \emph{K-Means} algorithm, and four training rounds for each dataset. For both encoding methods, We fix the number of zero coefficients. In the case of the LASSO, we regularize the hyper-parameter $\lambda = 1.2 / \sqrt[]{m}$, where $m$ is the window size. This assumption provides about ten nonzero coefficients \citep{mairal2009online}. In the case of OMP, we let the non-zero coefficients parameter be a $10\%$ of the number of atoms in the dictionary.

In the case of the Starlight dataset, we learn the dictionary using the training set and run evaluations in the test set. We also add  Gaussian noise to each light curve to train the dictionaries as real light curves. In the case of OGLE-3, we fold the light curves before building the dictionary through the standard procedure based on period calculation via the Lomb-Scargle algorithm \citep{lomb76,scargle1982studies}. We split the SMC with stratification into 4/5 of its samples to train dictionaries and 1/5 of its samples for the test the reconstruction.

\begin{figure}
  \centering
    \begin{subfigure}{0.5\textwidth}
        \includegraphics[width=0.95\textwidth,height=3.5cm,trim=0 15cm 0 0, clip=true]{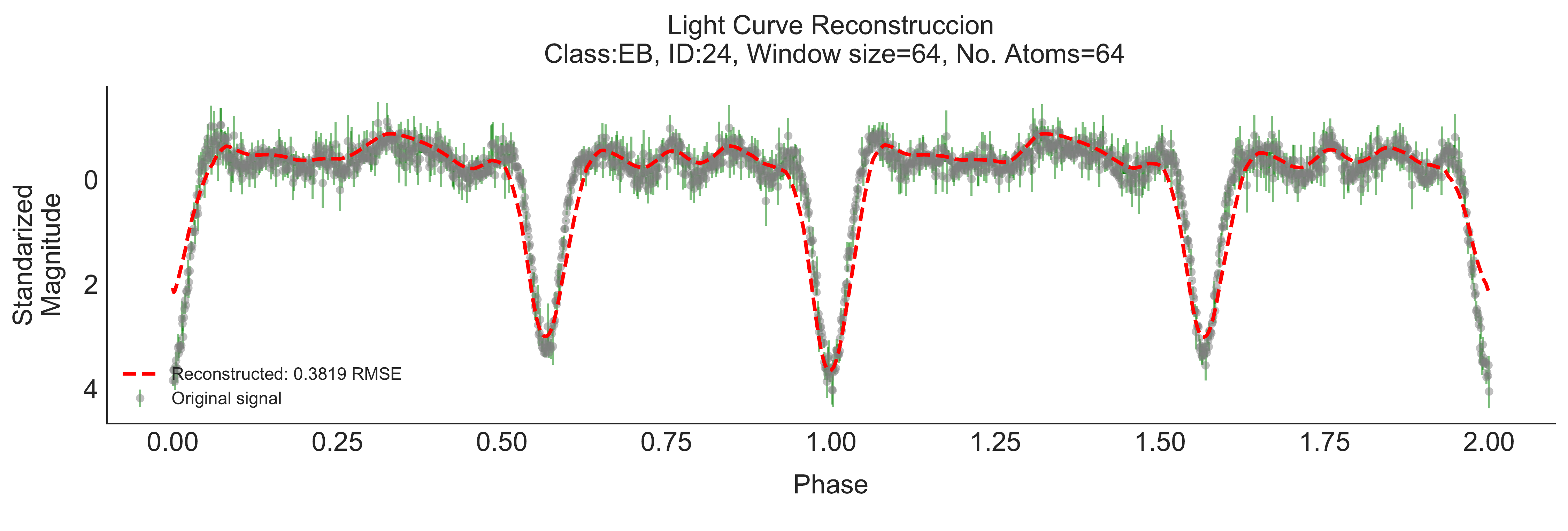}
        \caption{\label{ECLRecontructionSL-OMP}}
    \end{subfigure}
    \begin{subfigure}{0.5\textwidth}
        \includegraphics[width=0.95\textwidth,height=3.5cm,trim=0 15cm 0 0, clip=true]{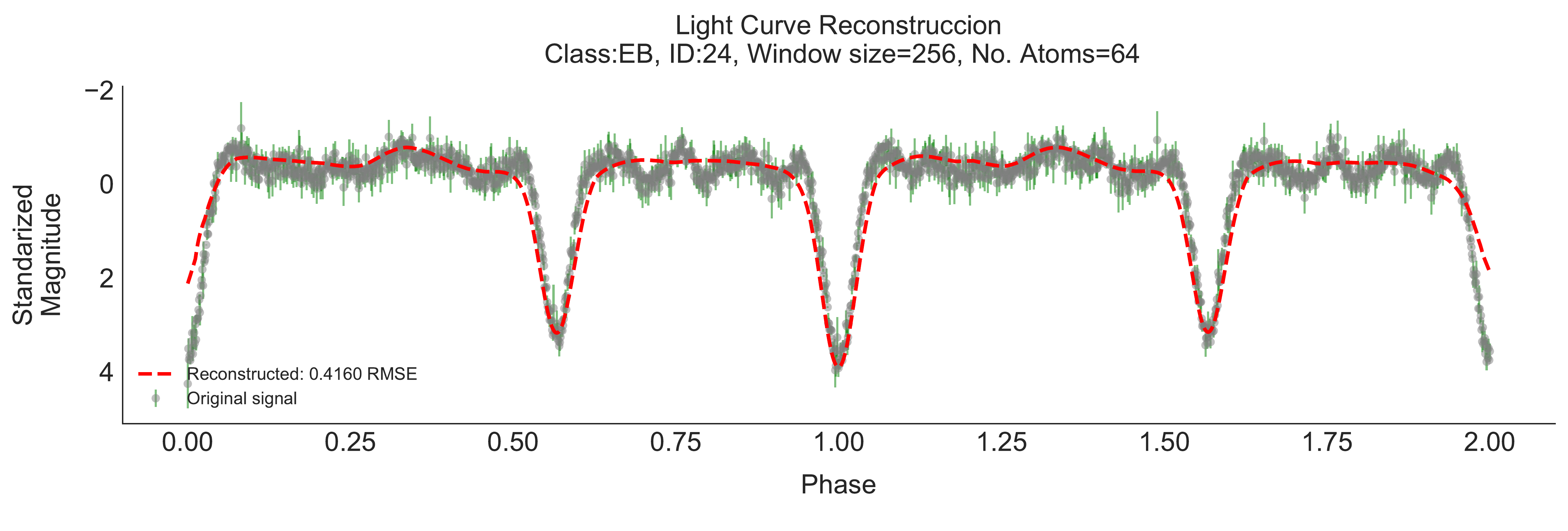}
        \caption{\label{RRLYRReconstructionSL-OMP}}
    \end{subfigure}
    \begin{subfigure}{0.5\textwidth}
        \includegraphics[width=0.95\textwidth,height=3.5cm,trim=0 15cm 0 0, clip=true]{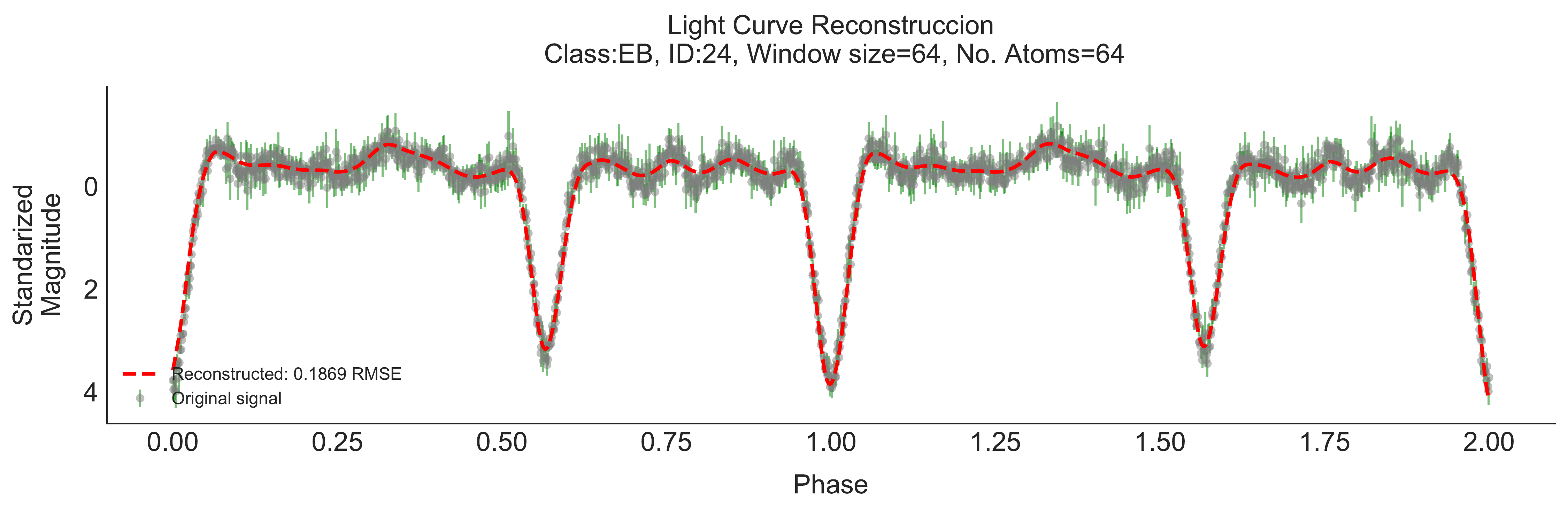}
        \caption{\label{ECLRecontructionSL-LASSO}}
    \end{subfigure}
    \begin{subfigure}{0.5\textwidth}
        \includegraphics[width=0.95\textwidth,height=3.5cm,trim=0 15cm 0 0, clip=true]{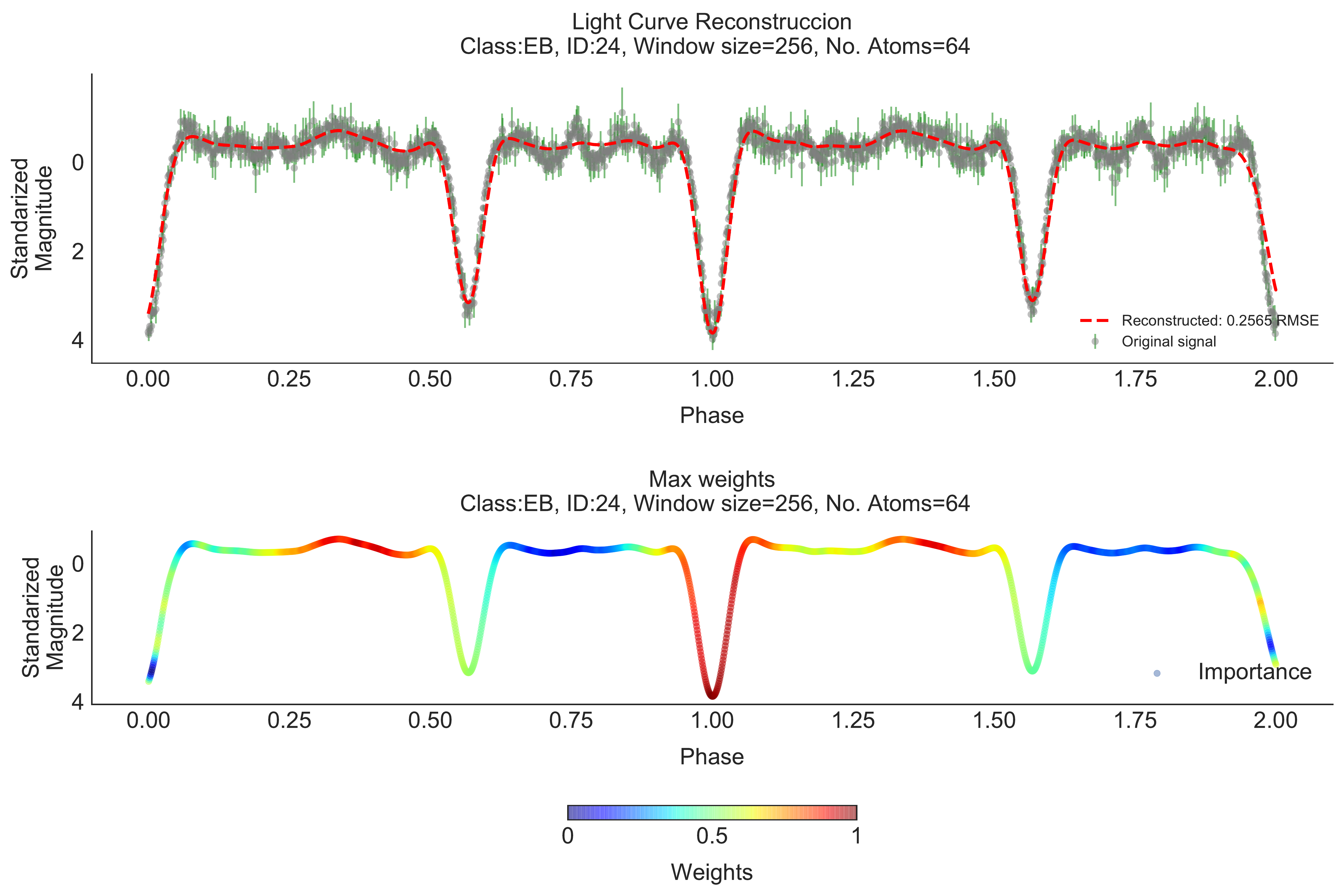}
        \caption{\label{RRLYRReconstructionSL-LASSO}}
    \end{subfigure}
    \caption{Examples of light curves reconstruction in the StarLight dataset using OMP and LASSO decoding algorithms. We observe that using smaller windows both algorithms reconstruct a more detailed version of the input. Longer windows provide a smoother version. The gray line is the original time series, and the red one denotes the reconstructed light curve. In all the examples we use a dictionary with 192 atoms. \subref{ECLRecontructionSL-OMP})~OMP decoding using a window size of 64. \subref{RRLYRReconstructionSL-OMP})~OMP decoding using a window size of 256. \subref{ECLRecontructionSL-LASSO})~LASSO decoding using a window size of 64. \subref{RRLYRReconstructionSL-LASSO})~LASSO decoding using a window size of 256.}
    \label{ReconstructionStarLight}
\end{figure}

\begin{figure}
  \centering
    \begin{subfigure}{0.5\textwidth}
        \includegraphics[width=0.95\textwidth,height=3.5cm,trim=0 15cm 0 0, clip=true]{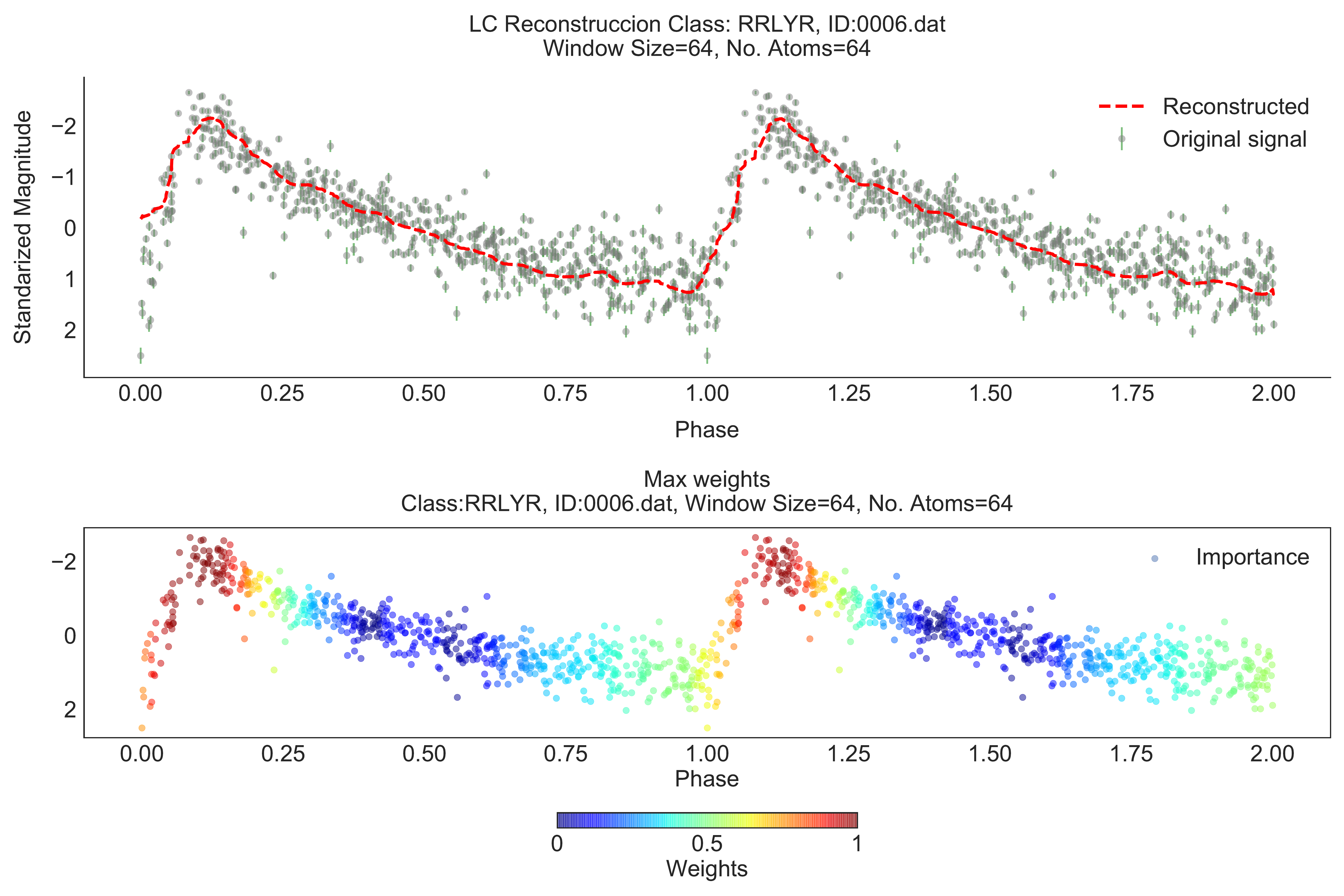}
        \caption{\label{RRLOGLEReconstructionOMP1}}
    \end{subfigure}
    \begin{subfigure}{0.5\textwidth}
        \includegraphics[width=0.95\textwidth,height=3.5cm,trim=0 15cm 0 0, clip=true]{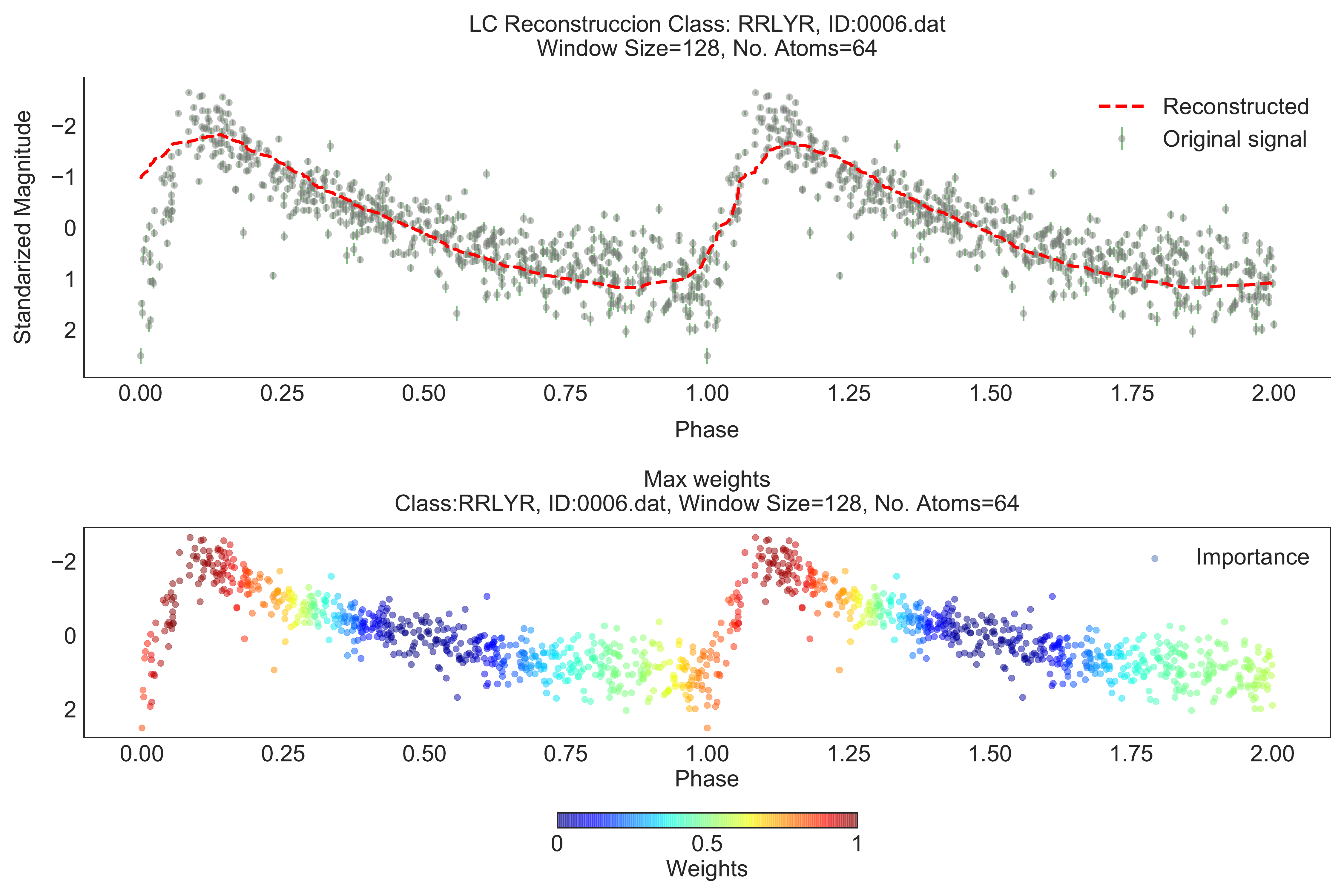}
        \caption{\label{RRLOGLEReconstructionOMP2}}
    \end{subfigure}
    \begin{subfigure}{0.5\textwidth}
        \includegraphics[width=0.95\textwidth,height=3.5cm,trim=0 15cm 0 0, clip=true]{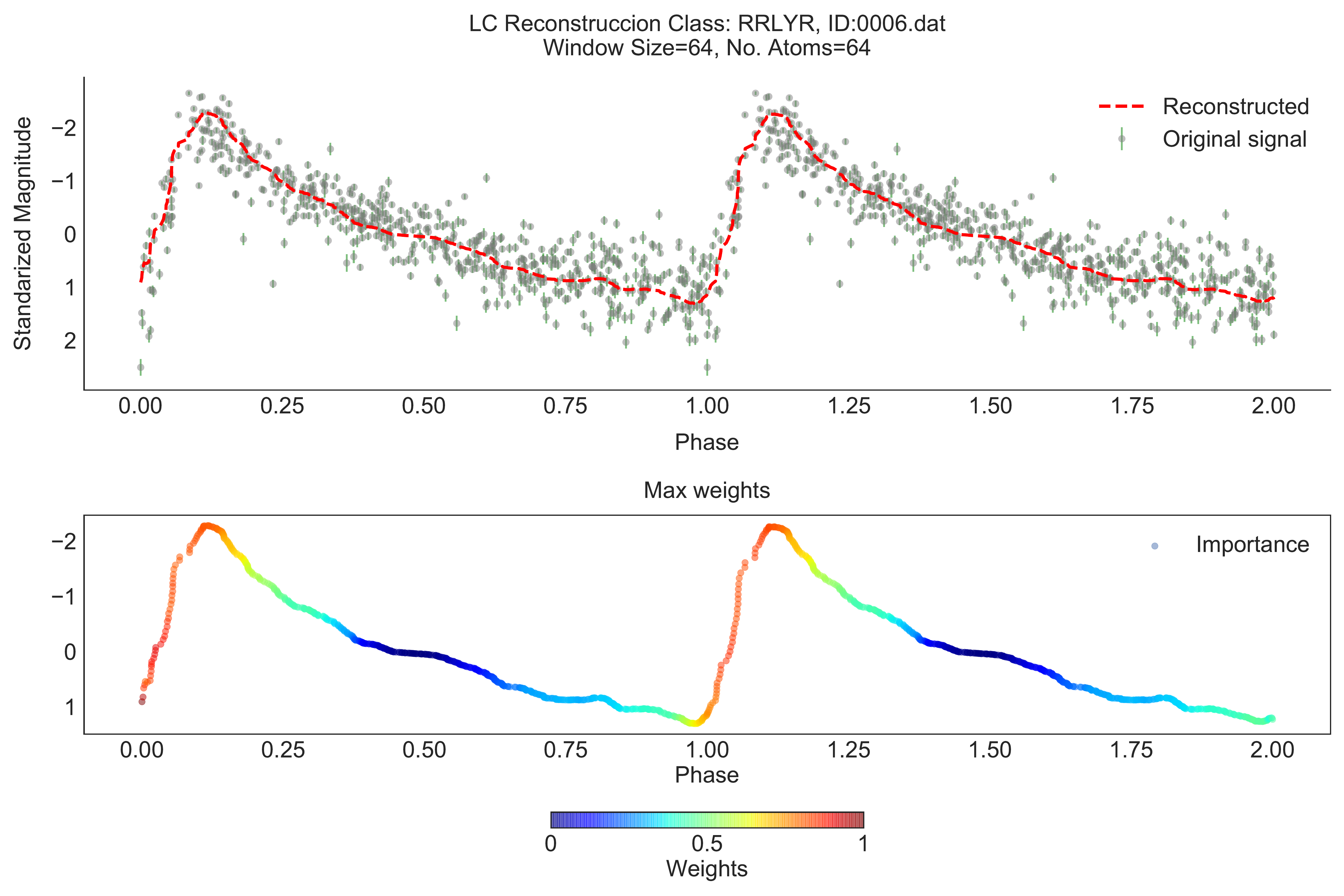}
        \caption{\label{RRLOGLEReconstructionLasso1}}
    \end{subfigure}
    \begin{subfigure}{0.5\textwidth}
        \includegraphics[width=0.95\textwidth,height=3.5cm,trim=0 15cm 0 0, clip=true]{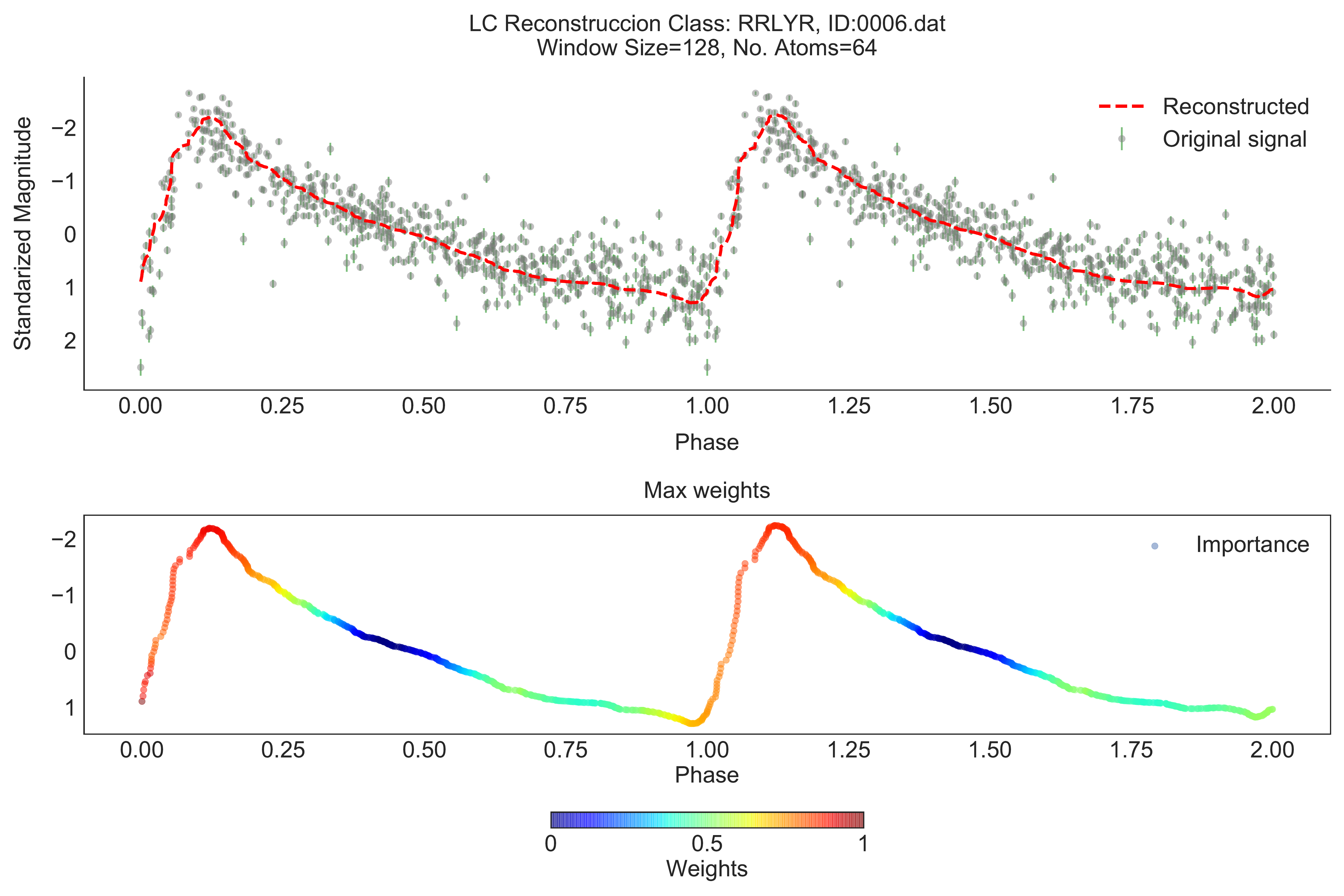}
        \caption{\label{RRLOGLEReconstructionLasso2}}
    \end{subfigure}
    \caption{Qualitative results of light curves reconstruction in the OGLE-3 dataset using OMP and LASSO decoding algorithms. Similarly that the StarLight dataset, we observe that smaller windows provide a more detailed version of the input signal. The gray line is the original time series, and the red one denotes the reconstructed light curve. In all the examples we use a dictionary with 192 atoms. \subref{RRLOGLEReconstructionOMP1})~OMP decoder using a window size of 64. \subref{RRLOGLEReconstructionOMP2})~OMP decoder using a window size of 128. \subref{RRLOGLEReconstructionLasso1})~LASSO decoder using a window size of 64. \subref{RRLOGLEReconstructionLasso2})~LASSO decoder using a window size of 128. The formulation of out framework does not consider errors related to measurements. Results provide a visualization with a smooth version of the original light curve.}
    \label{ReconstructionOGLE}
\end{figure}

Figures \ref{ReconstructionStarLight} and \ref{ReconstructionOGLE} show qualitative results for reconstruction on the StartLight and OGLE-3 datasets using OMP and LASSO as approximation method. These results point that the use of smaller windows during encoding allows for reconstructing a more detailed version of the input. Contrary, longer windows provide a smoother version of the approximation. Table \ref{tb:ReconstructionError} summarizes the results of applying the reconstruction algorithm on test light curves using different configurations of window size and atom numbers in OGLE-3. Results show that OMP achieves a low reconstruction error using the size of a window of $m=64$ and $64$ atoms, and LASSO using the size of a window of $m=384$ and $96$ atoms. A significance test based on the Kolmogorov-Smirnov framework points that there are no significant differences between the reconstruction errors provided by LASSO or OMP. However, in computation time OMP delivers faster results than LASSO. In the rest of the paper, we present results using OMP as the encoder.


\begin{table}
  \centering
  \caption{Reconstruction error. Each column shows the average of RMS error (RMSE) between the source light curve and its reconstructed version using our framework. We use this evaluation to select the best configuration. The p-value points that there is no significant difference in the reconstruction error between the two encoder methods. We use the Kolmogorov-Smirnov to run the significant test.}
  \label{tb:ReconstructionError}
    \begin{tabular}{rccr}
    \hline
    \multicolumn{1}{c}{\textbf{Configuration}} & \multicolumn{1}{p{4.645em}}{\textbf{RMSE LASSO}} & \multicolumn{1}{p{4.145em}}{\textbf{RMSE OMP}} & \multicolumn{1}{c}{\textbf{p-value}} \\
    \hline
    w:64 - a:32 & 0.899 & 0.907 & 1.6E-10 \\
    w:64 - a:64 & 0.887 & 0.885 & 4.7E-19 \\
    w:64 - a:96 & 0.878 & 0.889 & 1.7E-13 \\
    w:128 - a:32 & 0.900 & 0.928 & 7.2E-09 \\
    w:128 - a:64 & 0.891 & 0.930 & 5.9E-13 \\
    w:128 - a:96 & 0.858 & 0.919 & 7.4E-165 \\
    w:256 - a:32 & 0.900 & 0.963 & 1.6E-35 \\
    w:256 - a:64 & 0.869 & 0.938 & 9.9E-167 \\
    w:256 - a:96 & 0.829 & 0.940 & 0.0E+00 \\
    w:384 - a:32 & 0.905 & 0.993 & 9.4E-111 \\
    w:384 - a:64 & 0.859 & 0.974 & 0.0E+00 \\
    w:384 - a:96 & 0.803 & 0.970 & 0.0E+00 \\
    \end{tabular}%
  \label{tab:addlabel}%
\end{table}%

\begin{figure}
  \centering
    \begin{subfigure}{0.5\textwidth}
        \includegraphics[width=0.95\textwidth,trim=0cm 15cm 0cm 0cm,clip=true]{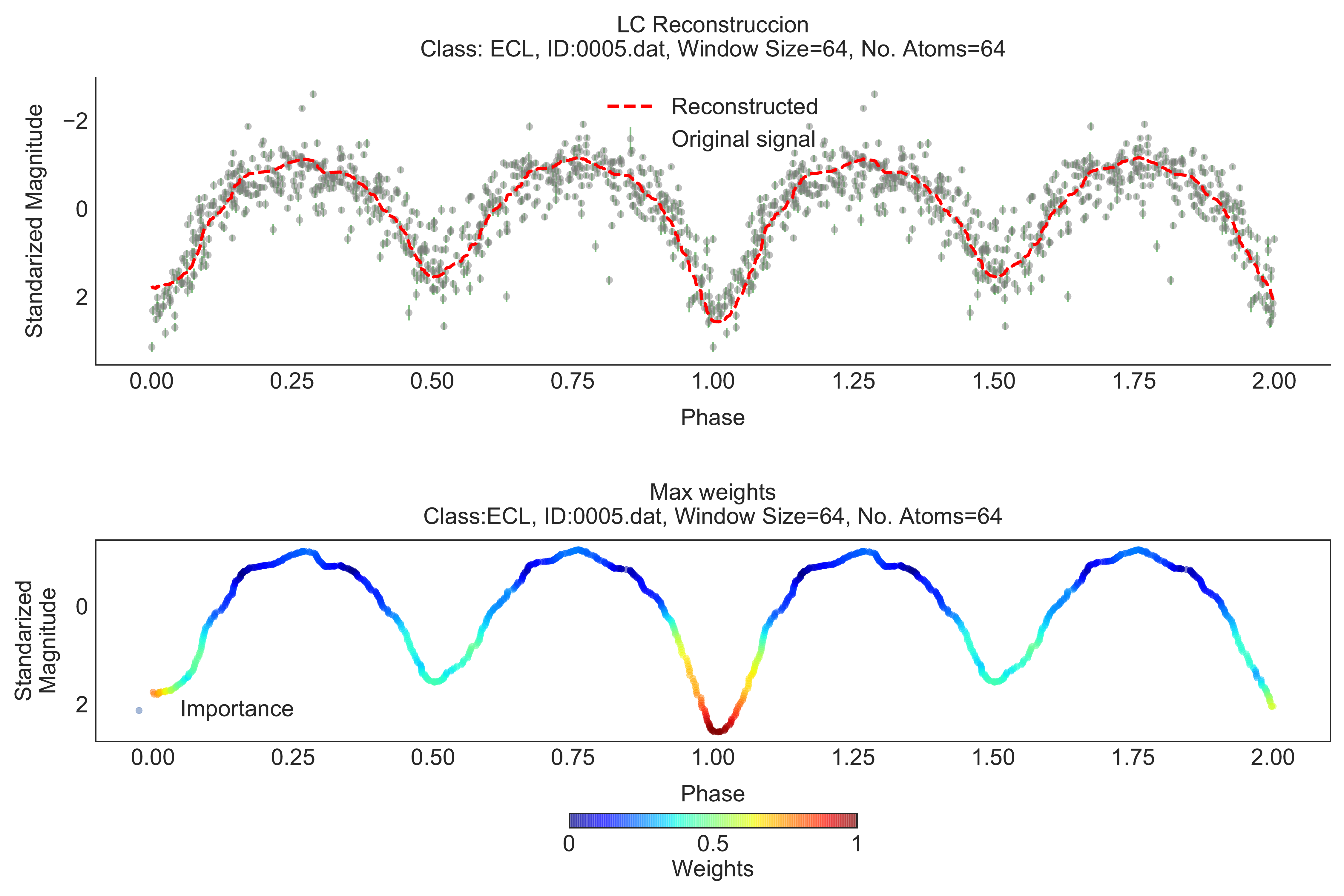}
        \caption{\label{ECLRecontruction}}
    \end{subfigure}
    \begin{subfigure}{0.5\textwidth}
        \includegraphics[width=0.95\textwidth,trim=0cm 15cm 0cm 0cm,clip=true]{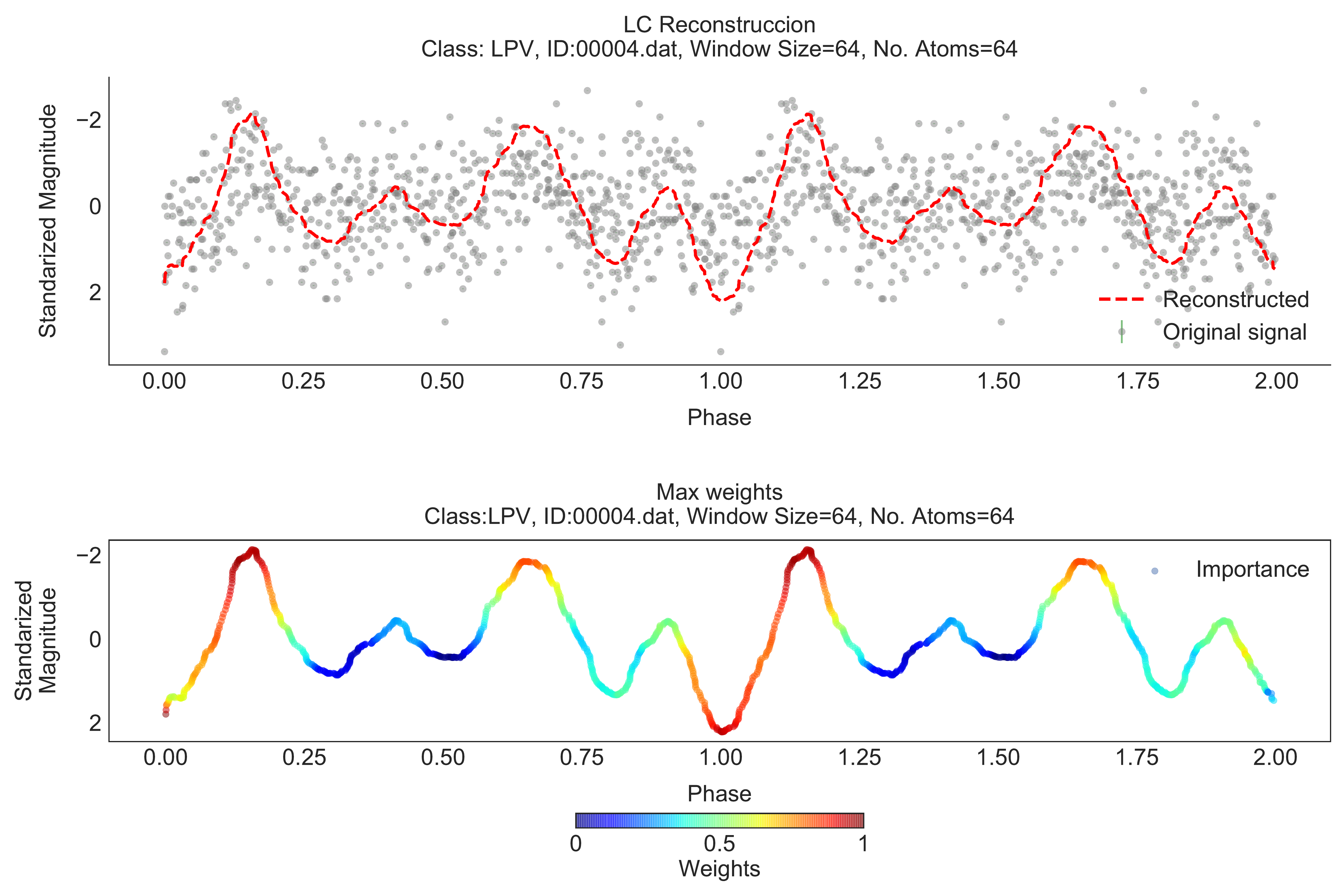}
        \caption{\label{LPVReconstruction}}
    \end{subfigure}
    \begin{subfigure}{0.5\textwidth}
        \includegraphics[width=0.95\textwidth,trim=0cm 15cm 0cm 0cm,clip=true]{./imgs/OGLE/OGLE-3_w64_320_omp.pdf}
        \caption{\label{RRLYRReconstruction}}
    \end{subfigure}
    \caption{Examples of light curves reconstruction. In all the examples we use the OMP method as the estimator. The gray line is the original time series, and the red one denotes the reconstructed light curve. Note that in all the cases the reconstructed light curve provides a more precise version of the original time series. \subref{ECLRecontruction}) ECL. \subref{LPVReconstruction}) LPV. \subref{RRLYRReconstruction}) RR-Lyrae.}
    \label{ReconstructionClasses}
\end{figure}

Figure \ref{ReconstructionClasses} shows three reconstruction examples of stars classes (ECL, LPV, and RRL) from the SMC set. Results show the reconstruction algorithm provides an overview of the original light curve. Furthermore, the output of our algorithm yields a smooth version of the input, removing most of the variations without fitting any regression model or filtering procedure. This effect is one of the advantages of sparse-based methods. In this way, we can visually distinguish the pattern of stars even for non-expert users.

\begin{figure}[htpb]
    \centering
    \begin{subfigure}{0.5\textwidth}
        \includegraphics[width=0.95\textwidth,height=6cm]{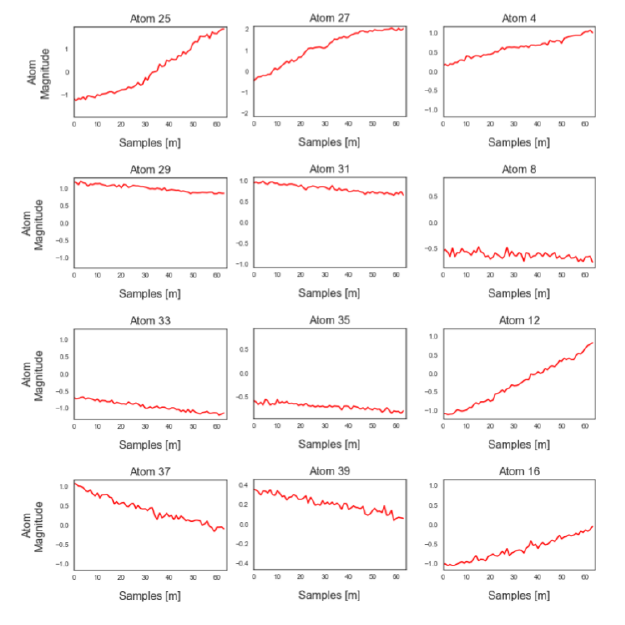}
        \caption{\label{Dictionary64}}
    \end{subfigure}
    \begin{subfigure}{0.5\textwidth}
        \includegraphics[width=0.95\textwidth,height=6cm]{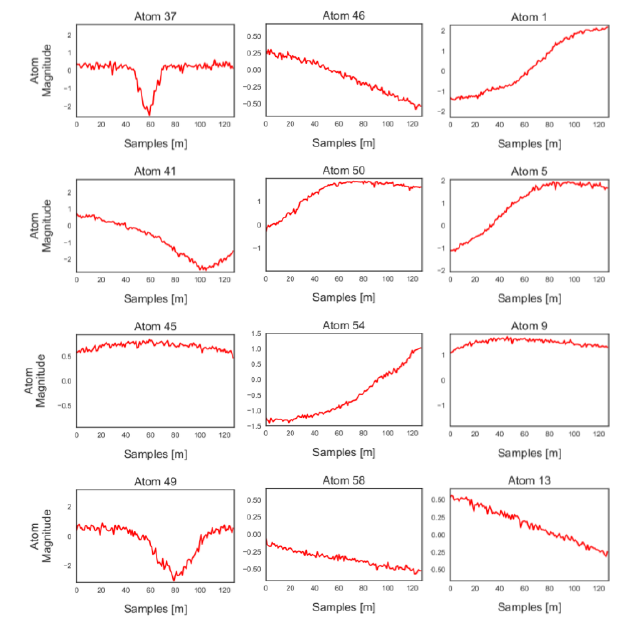}
        \caption{\label{Dictionary128}}
    \end{subfigure}
    \caption{Examples of atoms found during the dictionary training. We note a trade-off between the window size and distinguishing patterns relatives to each star class. \subref{Dictionary64}) Examples of a dictionary using windows size $m=64$. \subref{Dictionary64}) Examples of a dictionary using windows size $m=128$.}
    \label{AtomsExamples}
\end{figure}

Figure \ref{AtomsExamples} provides two examples of the atoms found by the algorithm used to build the supervised dictionary using different window sizes. Results show that the dictionary training found a set of atoms with high coherence with the original data. It is possible to note in some of the atoms the visual pattern related with each star class. In our results, we note that narrower window sizes provide an excellent approximation; however, in some cases, it is difficult to distinguish in the atom a pattern relative to each star. In the case of broader window size, we observe a less precise reconstruction, but at the same time, still recognizing a pattern.

\begin{figure}
    \centering
    \begin{subfigure}{0.5\textwidth}
        \includegraphics[width=0.95\textwidth,height=6cm]{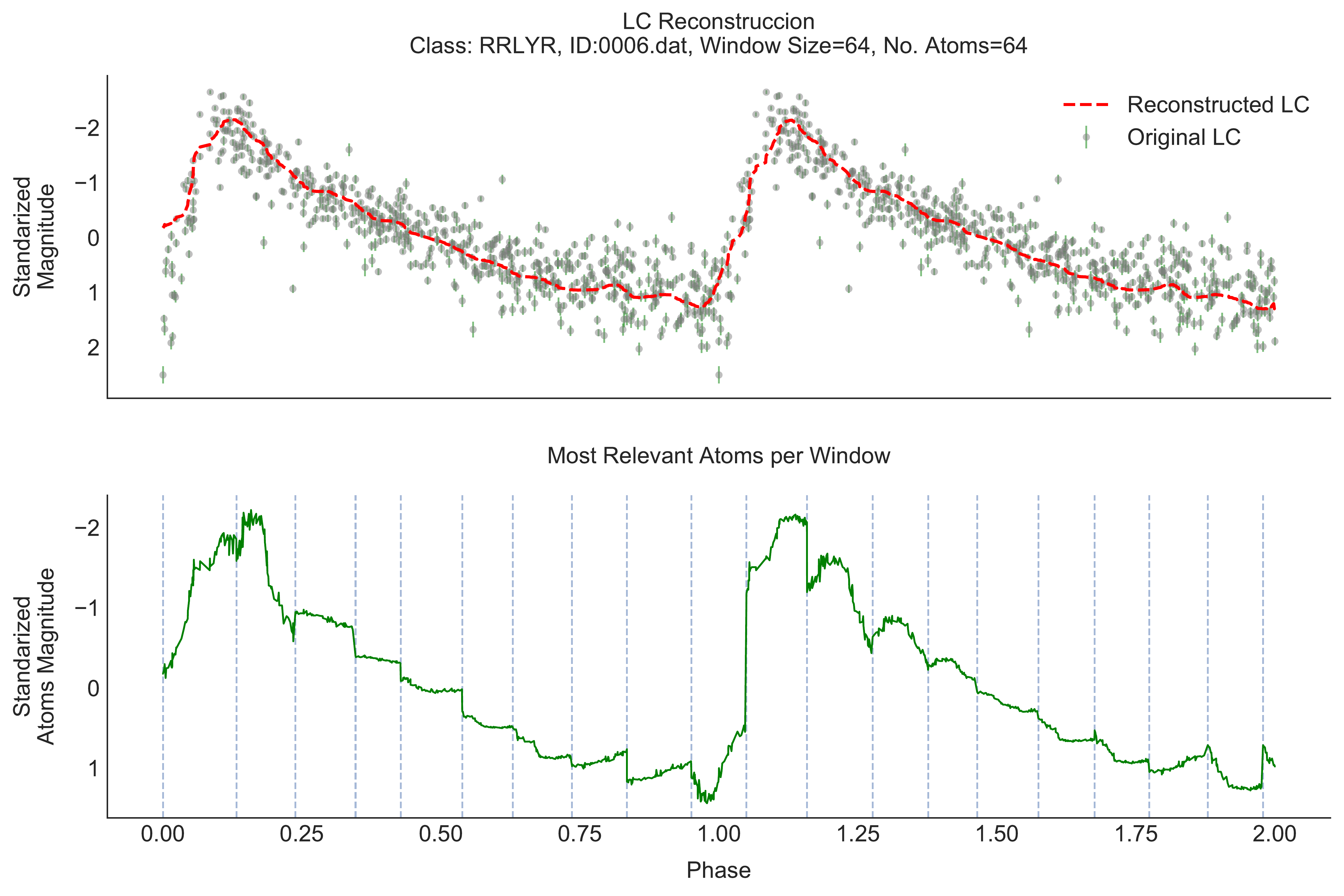}
        \caption{\label{UsingDictionary64}}
    \end{subfigure}
    \begin{subfigure}{0.5\textwidth}
        \includegraphics[width=0.95\textwidth,height=6cm]{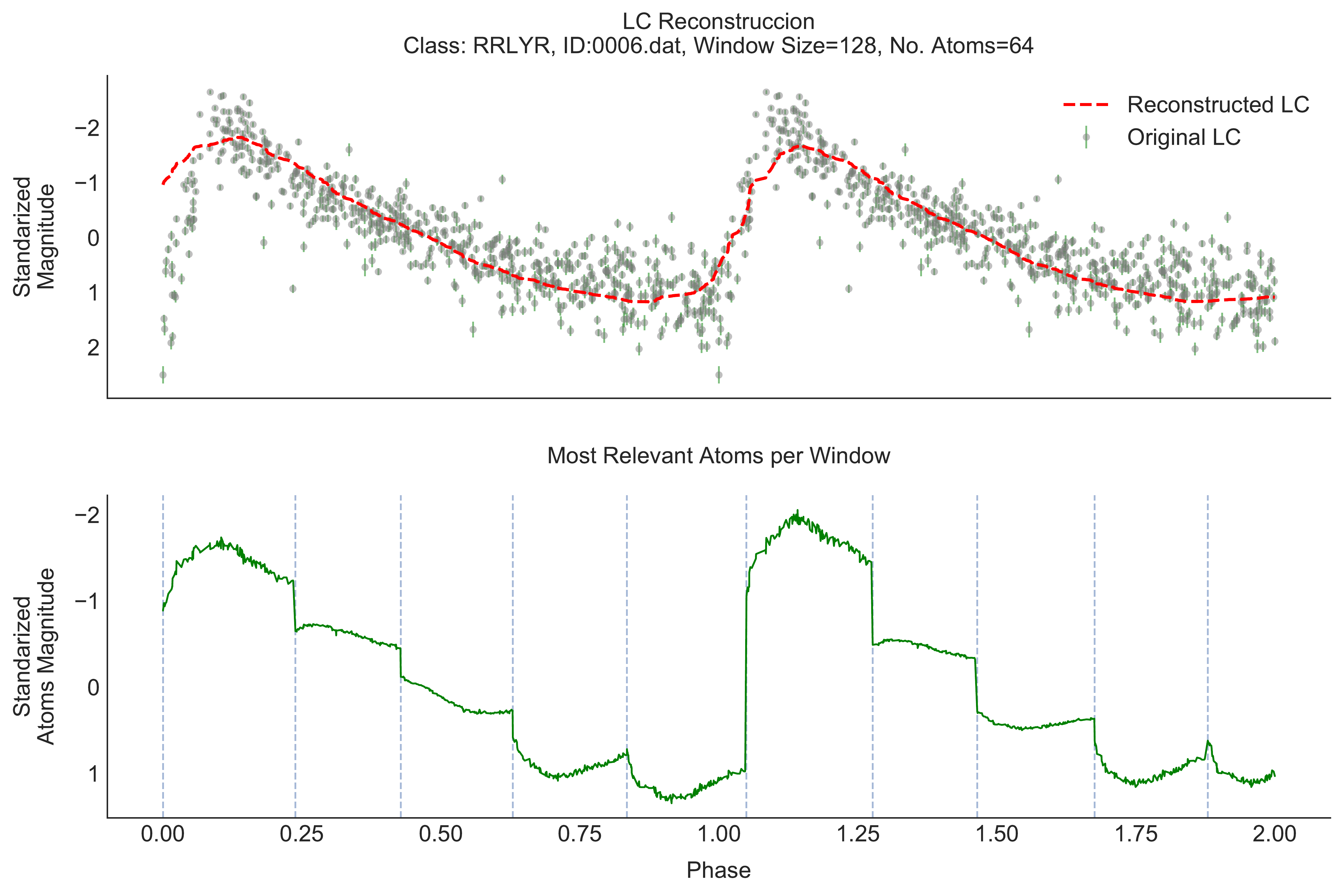}
        \caption{\label{UsingDictionary128}}
    \end{subfigure}
    \begin{subfigure}{0.5\textwidth}
        \includegraphics[width=0.95\textwidth,height=6cm]{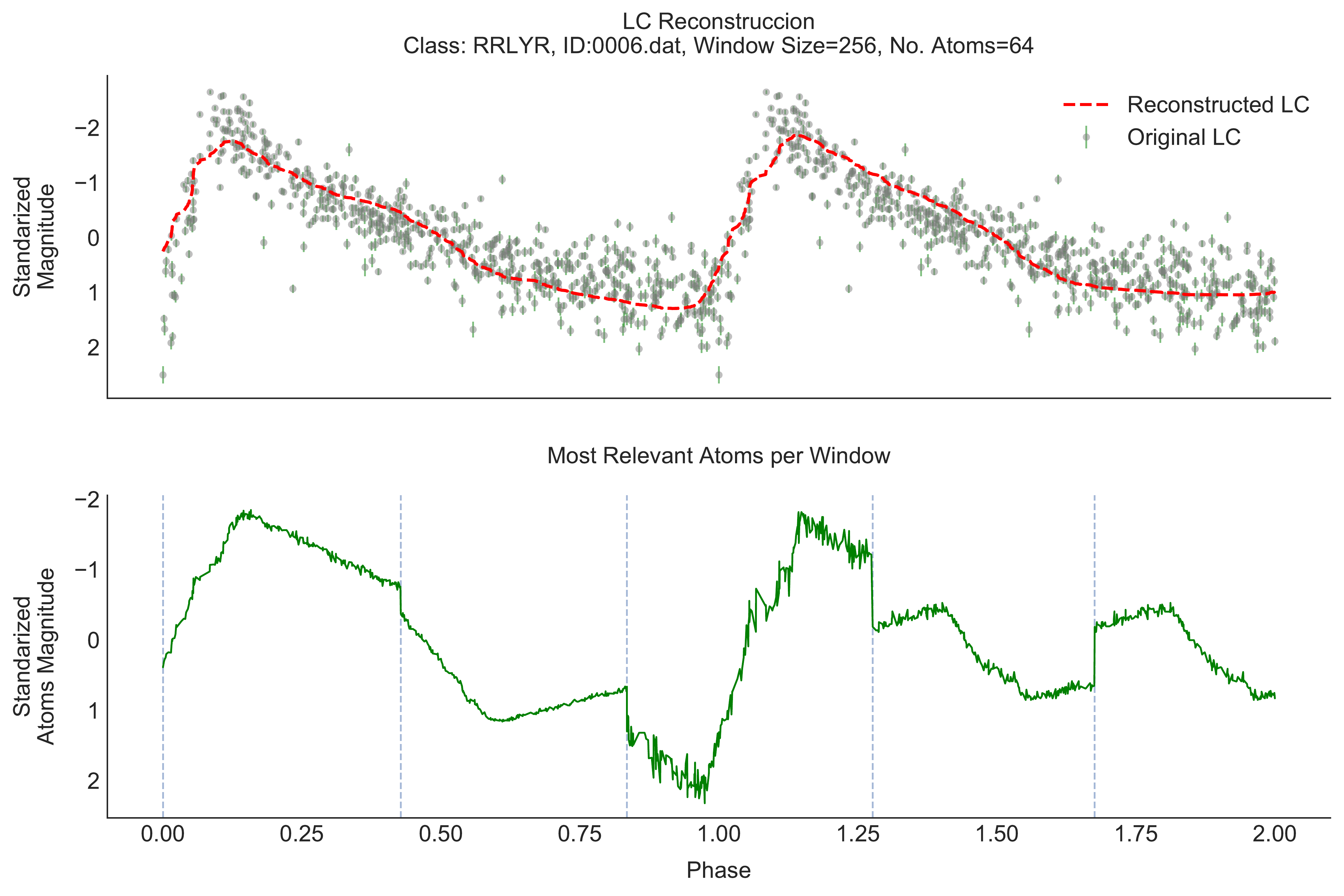}
        \caption{\label{UsingDictionary256}}
    \end{subfigure}
    \caption{Examples of relevant atoms found after running a rolling window of different sizes along the light curve. It is easy to note the trade-off between the window size used to run the light curve and the observed patterns relative to each star class. Although the abrupt changes between windows, the aims is to show a qualitative comparison. \subref{UsingDictionary64}) Using a windows size $m=64$. \subref{UsingDictionary128}) Using a windows size $m=128$. \subref{UsingDictionary128}) Using a windows size $m=256$.
    }
    \label{RelevantAtomsExamples}
\end{figure}

To clarify this point, we run non-overlapped rolling windows along a folded light curve. This visualization aims to show the trade-off in the selection of the window size. For each window, we get the atom with the maximum contribution to the reconstruction. Figure \ref{RelevantAtomsExamples} shows two examples of relevant atoms for different window sizes. We note that at a finner resolution most of the atoms do not have any particular similarity with the light curve. However, each part allows for an overall reconstruction and understanding of the data.

\subsection{Case Study 2: Highlighting Salient Parts}

\begin{figure}
  \centering
    \begin{subfigure}{0.5\textwidth}
        \includegraphics[width=0.95\textwidth,trim=0cm 3cm 0cm 15cm,clip=true]{./imgs/OGLE/ecl_OGLE-3_w64_320_omp.pdf}
        \caption{\label{ECLRecontruction}}
    \end{subfigure}
    \begin{subfigure}{0.5\textwidth}
        \includegraphics[width=0.95\textwidth,trim=0cm 3cm 0cm 15cm,clip=true]{./imgs/OGLE/lpv_OGLE-3_w64_320_omp.pdf}
        \caption{\label{LPVReconstruction}}
    \end{subfigure}
    \begin{subfigure}{0.5\textwidth}
        \includegraphics[width=0.95\textwidth,trim=0cm 0cm 0cm 15cm,clip=true]{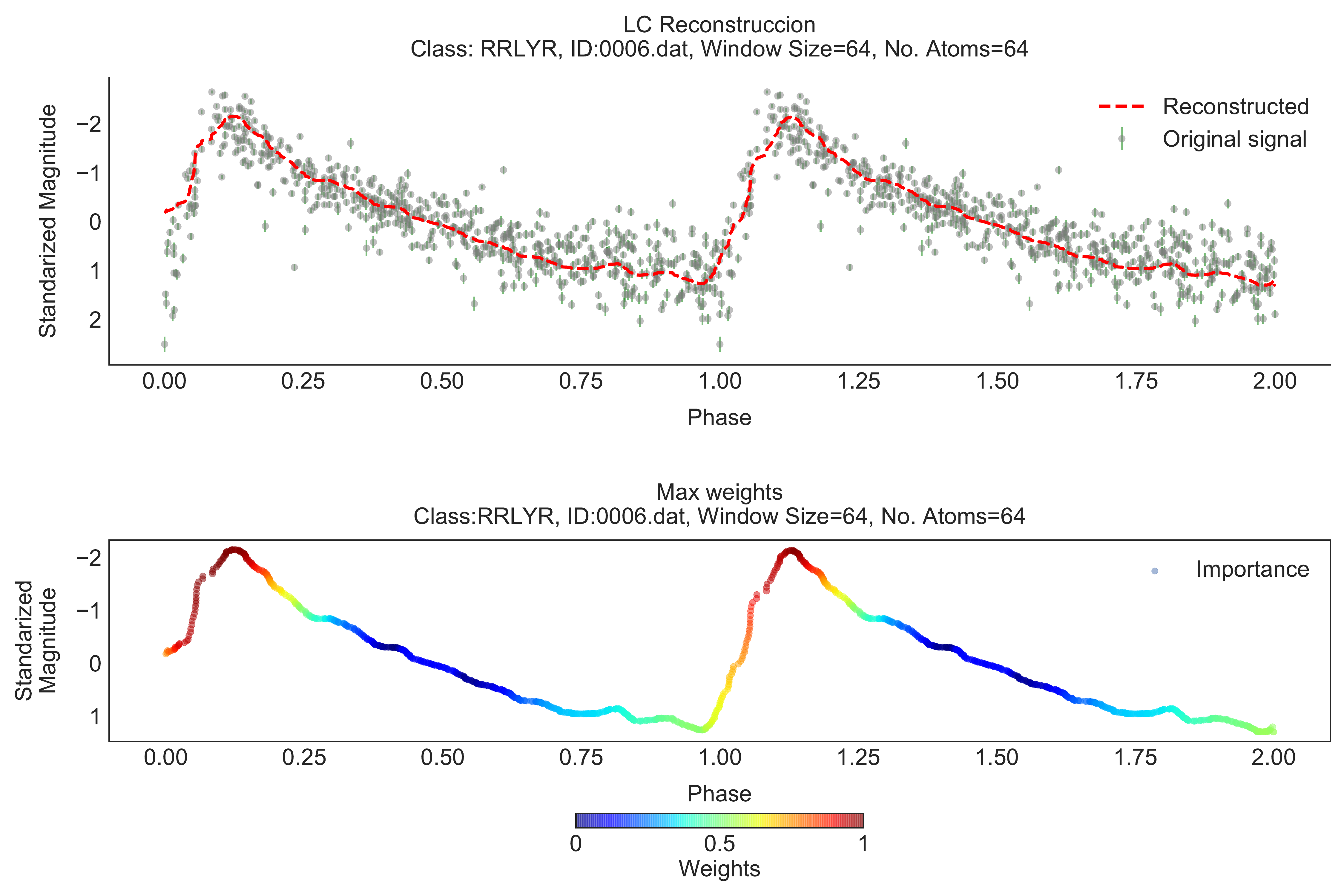}
        \caption{\label{RRLYRReconstruction}}
    \end{subfigure}
    \caption{Plots show examples of salient parts in the light curves provided by our method. Each figure is a heat map where red means a more salient part and blues a less relevant portion of the light curve. We plot the importance using the reconstructed versions. \subref{ECLRecontruction}) ECL. \subref{LPVReconstruction}) LPV. \subref{RRLYRReconstruction}) RR Lyrae.}
    \label{ResultsHeatMap}
\end{figure}

As we previously stated on Section \ref{sec:method}, our method uses the embedded information in the code vectors $\mathbf{\alpha}$ to recognize salient parts in time series. We apply our algorithm to encode and highlight these salient parts in real light curves in the I band from the OGLE-3 SMC data set. First, we create the dictionary using five variability classes. The algorithm learns the dictionary using 64 atoms per class giving a dictionary size of 320 atoms. We fix the windows size to $m=64$. Figure \ref{ResultsHeatMap} shows examples of the results obtained when coding light curves and then using these code vectors to highlight the salient parts in the time series. In red we can observe where the encoding put most of the effort during the approximation procedure. All of the examples present a tight relationship with the visual clues typically used by scientist to discriminate the variability class associated with a given light curve, which includes, for instance, the amplitude, maximum and minimum magnitudes, and rise time (i.e., the difference in phase between light curve maximum and minimum).

\section{Conclusion}
\label{sec:conclusion}

We have presented a sparse-based framework to find and highlight salient parts in astronomical light curves. Our algorithm takes advantage of the K-Means clustering to create a dictionary of atoms efficiently without memory overloads during training. Furthermore, this approach, combined with an encoding algorithm such as LASSO or OMP, provides us with a way to reconstruct the light curves. Both aspects of our approach can be useful when dealing with massive datasets.


We evaluated our method experimentally on synthetic and real-world datasets. In both cases, we found that our method produces high-quality reconstructions compared to the original version of these light curves. Moreover, based on the reconstruction, we provide an intuitive and visual representation of the salient parts in a light curve. By efficiently and automatically highlighting the light curve features that are most representative of a given variability class, this representation may help expert and non-expert users alike to guide the visual identification of stars in new catalogs. Besides, the identification of the sections of a light curve that are potentially most important may help in the planning of potentially time-consuming follow-up astronomical observations, by enabling the astronomer to focus her/his efforts mostly on these highlighted phases of variability. In the future, we would like to test the efficacy of the saliency delivered by our framework implementing a poll using our visualizations with experts.

\section*{Acknowledgements}

This research was funded by a grant for Interdisciplinary Research at Pontificia Universidad Cat\'olica de Chile (project PMI PUC1203). Additional support for this project is provided by the Ministry for the Economy, Development, and Tourism's Millennium Science Initiative through grant IC\,120009, awarded to the Millennium Institute of Astrophysics (MAS); by Proyecto Basal PFB-06/2007; by FONDECYT grant \#1171273; by CONICYT's PCI program through grant DPI20140066; and finally we acknowledge the support from CONICYT-Chile, through the FONDECYT Regular project number 1180054.




\bibliographystyle{mnras}
\bibliography{references}




%
%

\bsp	
\label{lastpage}
\end{document}